# Universal mobility characteristics of graphene originating from electron/hole scattering by ionised impurities

Jonathan H. Gosling,[1-2] Oleg Makarovsky,[1] Feiran Wang,[2] Nathan D Cottam,[1] Mark T. Greenaway,[3] Amalia Patanè,[1] Ricky Wildman,[2] Christopher J. Tuck,[2] Lyudmila Turyanska[2*] and T. Mark Fromhold[1*]

[1]School of Physics and Astronomy, University of Nottingham, Nottingham, NG7 2RD, UK.

[2]Centre for Additive Manufacturing, Faculty of Engineering, University of Nottingham, Nottingham, NG7 2RD, UK.

[3]Department of Physics, Loughborough University, Loughborough, UK.

Corresponding authors: Mark.Fromhold@nottingham.ac.uk

Lyudmila.Turyanska@nottingham.ac.uk

ABSTRACT

Pristine graphene and graphene-based heterostructures exhibit exceptionally high electron mobility and conductance if their surface contains few electron-scattering impurities. Here, we reveal a universal connection between graphene's carrier mobility and the variation of its electrical conductance with carrier density. Our model of graphene conductivity is based on a convolution of carrier density and its uncertainty, which reproduces the observed universality. Taking a single conductance measurement as input, this model accurately predicts the full shape of the conductance versus carrier density curves for a wide range of reported graphene samples. We verify the convolution model by numerically solving the Boltzmann transport equation to analyse in detail the effects of charged impurity scattering on carrier mobility. In this model, we also include optical phonons, which relax high-energy charge carriers for small impurity densities. Our numerical and analytical results both capture the universality observed in experiment and provide a way to estimate all key transport parameters of graphene devices.

Our results demonstrate how the carrier mobility can be predicted and controlled, thereby providing insights for engineering the properties of 2D materials and heterostructures.

1. INTRODUCTION

The unique electrical properties of graphene, such as high carrier mobility, $\mu > 10^4$ cm$^2$/Vs, at room temperature,[1-3] offer significant advantages for applications ranging from fast electronics to touch screens and ultrasensitive photon detection.[4-6] However, the emergence of graphene electronics on the market is limited by the absence of cost-effective large-scale production of high-quality graphene with reproducible electronic properties. The best results have been achieved in exfoliated suspended single-layer graphene (SLG) samples a few micrometres across, in which $\mu$ is limited only by the scattering of charge carriers (electrons and/or holes) by intrinsic phonons.[7,8] Epitaxial growth of graphene by CVD[9,10] or SiC-surface growth[11,12] methods provides cost-effective growth of large (>10 mm) SLG layers. However, the presence of charged impurities near the graphene significantly reduces $\mu$ due to the associated long-range Coulomb scattering centres.[7,13] If impurities are present, they often ionise and form charged scattering centres, which deflect the trajectories of electrons and holes in the 2D layer, thereby degrading the mobility. Charged impurities are introduced in the substrate and/or in the SLG-capping layer during the device processing, e.g. created by the diffusion of metallic ions present in the solvents or etching solutions used.[14]

Various theoretical models have been proposed to explain the effect of impurities on carrier mobility in SLG, and it is commonly accepted that $\mu$ is inversely proportional to the impurity density, $n_{imp}$, and independent of carrier density.[15,16] Scattering by charge-neutral point defects can also affect $\mu$, which is inversely proportional to the carrier density,[17,18] making it the dominant scattering mechanism at large carrier densities. The 2D nature of SLG means that it is sensitive to the surrounding environment, in particular the presence and position of the charged impurities. However, there is still limited understanding of the effect of the stand-off distance, $d$, of the impurities from the graphene plane on the carrier mobility, and other transport parameters. Additional complications arise in graphene-based heterostructures where SLG is sandwiched between two other materials with the same or different dimensionality (3D bulk, 2D layers and/or 0D quantum dots[2,19-21]).

Here, we report a theoretical and experimental study of the effects of charged impurities and optical phonons on the charge transport properties of graphene. We develop a graphene conductivity model based on the convolution of carrier density and its uncertainty. This model accurately reproduces the experimental data and provides a robust and simple way to model graphene conductivity as a function of the doping level and an applied gate voltage. To verify this model and extend the parameter and sample range investigated, we also present numerical *k*-space simulations of carrier transport based on a self-consistent model for the minimum carrier concentration and a full time evolution of the momentum distribution of the charge carriers. By using a Discontinuous Galerkin technique to solve the Boltzmann transport equation numerically, we analyse the effects of charged impurity scattering on the electron/hole mobility and show how such processes give rise to the universal behaviour that we have identified. We model the conductivity over a wide range of carrier densities in the presence of multiple sources of scattering and demonstrate that the mobility increases inversely as the conductance peak narrows. The calculations are supported by analysing experimental results obtained on both pristine and surface-decorated graphene devices. Our studies, for the first time, enable the development of quantitatively accurate, first-principles, calculations of the entire set of transport parameters: conductivity, mobility, carrier concentration and its uncertainty, in graphene layers containing charged impurities. These parameters are of fundamental importance to the full range of graphene-based heterostructures developed for 2D electronics.

## 2. THEORETICAL MODEL

In our work, we consider graphene sheets with charged impurities at a distance, $d_{imp}$, from the plane of the graphene and optical phonons, $\hbar\omega$ (Figure 1a). We model the effect of these two scattering mechanisms on the following electrical properties of graphene: the carrier concentration ($n$), mobility ($\mu$), conductivity ($\sigma$) and resistivity ($\rho$) at the Dirac point ($\rho_{max}$). Graphene conductivity in the vicinity of the Dirac point can be strongly affected by a number of different phenomena besides impurity scattering, including ballistic transport effects,[22] quantum capacitance[7,23] and temperature.[24,25] As a result, the device conductivity, and carrier density are non-zero even when the Fermi energy, $\varepsilon_F$, is at the Dirac (charge neutrality) point. Therefore, a simple model for the Drude conductivity:

$$\sigma_D = e\mu n_c, \qquad (1)$$

with a constant mobility, $\mu$, is not applicable for small gate voltages, assuming the constant capacitance model for graphene's carrier number density

$$n_c = |CV_g/e + n_0|, \qquad (2)$$

where $n_0 = n(V_g = 0)$ is the sheet density of the graphene doping (Supplementary Information, SI1). Spatial fluctuations of the local electrostatic potential in the graphene layer, and the presence of electron and hole 'puddles' (Figure 1b) are thought to explain the non-zero conductivity and resistance ($\sigma_{min}$ and $\rho_{max}$ Figure 1c-d) observed at the Dirac point.[15,26] Electrons and holes play equal roles in determining the graphene conductivity with no scattering at the borders between the n- and p-type graphene areas due to the Klein paradox.[27]

We now introduce a model that includes the effect of spatial fluctuations on the carrier number density versus gate voltage relation. Combined with the Drude conductivity, this model accurately describes the shape of the measured $\rho(V_g)$ curve (Figure 1d).

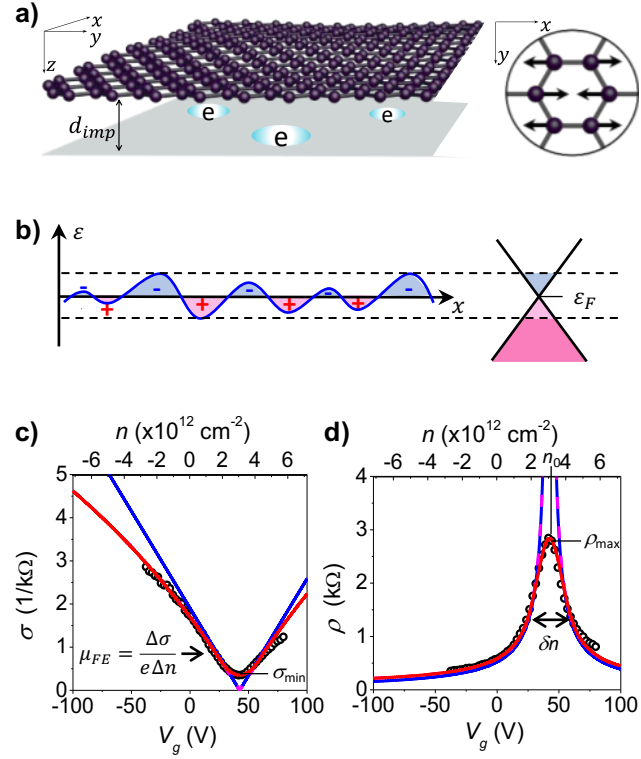

**Figure 1 (a)** Schematic diagram showing the impurity position with respect to the graphene sheet and a pictorial representation of spatio-temporal (angular frequency ω) phonon oscillations over one unit cell. **(b)** Spatial inhomogeneity in the impurity potential (left) and the resulting broadening of the energy distribution of electrons and holes near the Fermi level (Dirac cones, right). Conductivity **(c)** and resistivity **(d)** of graphene plotted versus gate voltage; circles show experimental data, blue lines show linear fits, using Equation (1), dashed pink curves are fits to the data found by including a short-range scattering resistivity, red curves result from the convolution fit.

To first approximation, the carrier number density at a given gate voltage, $n(V_g)$ can be modelled as the moving average (convolution) of $n_c$ [Equation (2)] over a window of width $\delta n$, which is the characteristic amplitude of the carrier density fluctuations in the graphene layer. Such a moving average is equivalent to the convolution of the $n_c(V_g)$ function with a box function $f(n_c)$ of width $\delta n$, see Supplementary Information, SI1, which gives

$$n = n_c * f(n_c), \text{ where}$$

$$f(n_c) = \begin{cases} \dfrac{1}{\delta n} & \text{for } -\dfrac{\delta n}{2} < n_c < +\dfrac{\delta n}{2} \\ 0 & \text{for } -\dfrac{\delta n}{2} > n_c \text{ or } n_c > +\dfrac{\delta n}{2} \end{cases}.$$

Using the linear $n_c(V_g)$ dependence in Equation (2) gives

$$n(V_g) = \begin{cases} \frac{\delta n}{4} + \frac{n_c^2(V_g)}{\delta n} & \text{for } n_c(V_g) < \frac{\delta n}{2} \\ n_c(V_g) & \text{for } n_c(V_g) > \frac{\delta n}{2} \end{cases}. \quad (3)$$

This expression for $n(V_g)$ is equal to the constant capacitance model for gate voltages where $n_c(V_g) > \frac{\delta n}{2}$ and has a parabolic form for gate voltages close to the Dirac point, where $n_c(V_g) < \frac{\delta n}{2}$. Using Equation (3), we can determine the conductivity,

$$\sigma_f(V_g) = e\mu n(V_g). \quad (4)$$

From this expression, it can be shown that $\delta n$ equals the full width half maximum of the peak resistivity around the Dirac point. Also, when $n_c = 0$, $\delta n = 4 n_{NP}$, where $n_{NP} = \sigma_{min}/e\mu$, is the residual carrier density at the Dirac (neutrality) point. Our model thus allows us to extend linear conductivity models[8] to a wider range of $n$ values around, and including, the Dirac point, thereby accurately reproducing the full observed $\sigma(V_g)$ and $\rho(V_g)$ dependences (Figure 1c-d).

To explain the observed $\delta n(V_g)$ and $\mu(V_g)$ curves in more detail, we now develop a semi-classical k-space simulations of the electron and hole dynamics. Graphene has a linear energy-wavevector dispersion relation, $\varepsilon = \pm \hbar v |\mathbf{k}|$, for the electrons and holes at the two inequivalent valleys, $\mathbf{K}$ and $\mathbf{K'}$ in reciprocal space. This results in a constant speed of the charge carriers, $v = 10^6$ ms$^{-1}$. These charge carriers undergo diffusive scattering transport, which we describe using a semi-classical Boltzmann transport approach. The influence of perturbations, such as impurities and phonons, on the scattering of electrons is calculated using the Fermi golden rule for transition rates between states. The electrons are initially assumed to obey a Fermi-Dirac distribution, $f_0(\mathbf{k})$. Inter-band transitions are neglected such that the valence band is assumed to be full throughout the time-evolution when the gate voltage is positive, i.e. when the chemical potential lies within the conduction band.

The spatially homogeneous Boltzmann Transport Equation,

$$\frac{\partial f(t,\bm{k})}{\partial t} = -\frac{e}{\hbar}\bm{E}\cdot\nabla_{\bm{k}}f(t,\bm{k}) + \left(\frac{\partial f(t,\bm{k})}{\partial t}\right)_{coll}, \qquad (6)$$

describes the evolution of the occupancy, $f(t,\bm{k})$, of state $\bm{k}$ at time $t$. The first term on the right-hand side of Equation (6) describes the acceleration of electrons under an applied electric field, $\bm{E}$, and the collision term is given by the detailed balance equation,

$$\left(\frac{\partial f(t,\bm{k})}{\partial t}\right)_{coll} = \sum_{\bm{k}'}[S_{\bm{k}'\to\bm{k}}f(t,\bm{k}')(1-f(t,\bm{k})) - S_{\bm{k}\to\bm{k}'}f(t,\bm{k})(1-f(t,\bm{k}'))], \qquad (7)$$

where $S_{\bm{k}\to\bm{k}'}$ is the transition rate of carriers from a state of crystal momentum $\hbar\bm{k}$ to a new state with momentum $\hbar\bm{k}'$. We solve Equation (6) using a Discontinuous Galerkin (DG) numerical simulation[28] for the steady-state distribution function, $f(\bm{k})$. We can then determine the mobility, $\mu = v_d/E$, for an applied electric field, $E$, and the resulting drift velocity,

$$v_d = \frac{1}{n\pi^2\hbar}\int f(\bm{k})\nabla_{\bm{k}}\varepsilon(\bm{k})\mathrm{d}\bm{k}. \qquad (8)$$

Alternatively, to find approximate analytical solutions to Equation (3) we can assume a small shift in the initial distribution function, $f_0(\bm{k})$, proportional to the ensemble momentum relaxation time, $\tau_m$. This results in the linearised Boltzmann approximation[29] for the mobility which, at zero-temperature, is related to the relaxation time at the chemical potential, $\varepsilon_F$, via

$$\mu = e\tau_m(\varepsilon_F)v^2/\varepsilon_F. \qquad (9)$$

We calculate the momentum relaxation time, $\tau_m(\bm{k})$, as the sum over all possible transition rates, $S_{\bm{k}\to\bm{k}'}$, modified by the deflection angle, $\theta_{\bm{k},\bm{k}'}$, between the incoming and outgoing vectors:

$$\frac{1}{\tau_m(\bm{k})} = \sum_{\bm{k}'}S_{\bm{k}\to\bm{k}'}\left(1-\cos\theta_{\bm{k},\bm{k}'}\right) \approx \frac{A}{(2\pi)^2}\int S_{\bm{k}\to\bm{k}'}\left(1-\cos\theta_{\bm{k},\bm{k}'}\right)\mathrm{d}\bm{k}', \qquad (10)$$

where $A$ is the area of graphene unit cell.

The effect of screening by the electron and hole gases is included by introducing a random phase approximation for the dielectric screening function[30,31]

$$\epsilon_{sc} = 1 - \tilde{v}_{2D}\Pi(q), \qquad (11)$$

where $\tilde{v}_{2D} = e^2/2\epsilon_0\epsilon_r q$ is the unscreened Coulomb potential in Fourier space, $\Pi(q)$ is the static polarisation function and the reciprocal space variable $q = |\mathbf{k'}-\mathbf{k}|$. Since the conduction band distribution function changes throughout the simulation, the screening function should be carefully considered. However, the valence band distribution is constant throughout as the band is assumed to be full. The polarisation function for screening by a full valence band is[31]

$$\Pi_{val} = -\frac{q}{4\hbar v}. \tag{12}$$

For the conduction band, maximum screening in the Thomas-Fermi limit, $q \to 0$, occurs at $T=0$ and is given by

$$\Pi_{TF} = \int D(\varepsilon)\frac{\partial f}{\partial \varepsilon}d\varepsilon = -\frac{2}{\pi(\hbar v)^2}\int f d\varepsilon = -\frac{2}{\pi(\hbar v)^2}\varepsilon_F, \tag{13}$$

where $D(\varepsilon) = 2\varepsilon/\pi(\hbar v)^2$ is the density of states. Throughout the simulation, the integral in Equation (10) does not change, due to conservation of charge. Since both Equations (12) and (13) are independent of the evolution of the distribution function, we define a time-independent two-regime screening function, where Thomas-Fermi screening is assumed for low-energy scattering and the valence electron screening is assumed for high-energy electrons[15], i.e. we set

$$\epsilon_{sc} = \begin{cases} 1 + \frac{q_s}{q} & \text{for } q \leq \frac{8}{\pi}k_F \\ 1 + \frac{\pi r_s}{2} & \text{for } q > \frac{8}{\pi}k_F \end{cases}, \tag{14}$$

where $r_s = e^2/(4\pi\epsilon_0\epsilon_r\hbar v)$ and $q_s = 4k_F r_s$. For graphene on $SiO_2$, we take $\epsilon_r \approx 2.45$ [16].

A Coulombic scattering potential is assumed for charged impurities near the graphene plane:

$$U(r) = \frac{e^2}{4\pi\epsilon_0\epsilon_r\sqrt{r^2+d_{imp}^2}}, \tag{15}$$

where $d_{imp}$ is the distance of the impurities from the graphene plane. Note that in this model we consider randomly-distributed impurities and disregard any possible spatial correlation of charges below and above the graphene plane[32,33]. Then the transition rate is

$$S^{imp}_{\mathbf{k}\to\mathbf{k}'} = n_0 \frac{\pi}{A\hbar} \left|\frac{2\pi e^2 e^{-qd_{imp}}}{\kappa q \epsilon_{sc}(q)}\right|^2 (1 + \cos\theta_{\mathbf{k},\mathbf{k}'})\delta(\varepsilon_{\mathbf{k}'} - \varepsilon_{\mathbf{k}}), \tag{16}$$

where $q = 2k\sin(\theta_{\mathbf{k},\mathbf{k}'}/2)$ for elastic scattering. Defects that perturb the band structure over a small spatial area are characterised by the short-range scattering potential $U(r) = U_0 H(R-r)$, where $H$ is the Heaviside step function and $R$ gives the spatial extent of the perturbation. This potential represents any charge-neutral defects within the lattice. The rate of carrier scattering transitions due to such defects is

$$S^{sr}_{\mathbf{k}\to\mathbf{k}'} = n_{sr} \frac{\pi}{A\hbar} \left|\frac{\sigma_c U_0}{\epsilon_{sc}(q)}\right|^2 (1 + \cos\theta_{\mathbf{k},\mathbf{k}'})\delta(\varepsilon_{\mathbf{k}'} - \varepsilon_{\mathbf{k}}), \tag{17}$$

where $\sigma_c = \pi R^2$ is the effective cross-section of defects with an areal density $n_{sr}$.

Although phonons are not expected to have a significant effect on the mobility of SLG compared to that of the impurities[7], we include optical phonons in our simulation because, for small impurity densities, carriers can be accelerated to sufficiently high energies to emit an optical phonon before they are scattered by impurities. We also consider the low temperature regime ($k_B T \ll \hbar\omega$) and assume that the phonon occupation is $N = 0$, such that only spontaneous emission occurs. We use the optical phonon scattering rates calculated using density functional theory in Refs. [34] and [35]. Near the $\mathbf{\Gamma}-$ points of the reciprocal lattice, the energy and coupling strength of both transverse and longitudinal optical phonons are reported to be $\hbar\omega_O \approx 165$ meV and $\beta_O \approx 10$ eV/Å, respectively[36]. Therefore, we can combine the transition rates of the two modes to obtain a single overall optical scattering rate

$$S^O_{\mathbf{k}\to\mathbf{k}'} = \frac{2\pi\beta_O^2}{A\rho\omega_O}\delta(\varepsilon_{\mathbf{k}'} - \varepsilon_{\mathbf{k}} + \hbar\omega_O). \tag{18}$$

Phonons at the $\mathbf{K}-$ points cause intervalley scattering at a rate

$$S^K_{\mathbf{k}\to\mathbf{k}'} = \frac{2\pi\beta_K^2}{A\rho\omega_K}(1 - \cos\theta_{\mathbf{k},\mathbf{k}'})\delta(\varepsilon_{\mathbf{k}'} - \varepsilon_{\mathbf{k}} + \hbar\omega_K), \tag{19}$$

where $\hbar\omega_K \approx 124$ meV and $\beta_K \approx 3.5$ eV/Å.[36]

We consider a residual charge density of electron-hole puddles at the Dirac point, due to inhomogeneity in the impurity-induced potential (Figure 1b), which limits the minimum conductivity. This residual charge density was calculated in Ref. [15] assuming a random distribution of impurities. Here, we assume that the transition from the residual charge-dominated minimum carrier concentration to the linearly $V_g$-dependent concentration occurs when the gate-induced charged density, $n(V_g – V_0)$, is equal to the residual charge density, $n_{NP}$, where $V_0$ is the position of the Dirac point.

For all numerical simulations, we apply an electric field, $E = 10^4$ V/m (0.1 V drop across a 10 µm long SLG), corresponding to a regime of low-field mobility, where $\mu$ is independent of the applied electric field strength. For comparison, we also calculate the mobility using the linearised Boltzmann formalism, Equation (9), with the scattering time calculated using Equation (10), in which the integrals are evaluated numerically. The linearised Boltzmann method is simpler, but the DG method is more accurate and allows consideration of the form of the ensemble momentum distribution, which we will show is important for small impurity densities where electrons can be accelerated to high energies, far from thermal equilibrium.

Figure 2a shows the calculated dependence of mobility on $n$ for $n > n_{NP}$. We define the low-carrier mobility regime as when the chemical potential is near the Dirac point. With increasing $n$, we observe an initial increase of $\mu$, as it is dominated by impurity scattering. For larger $n$, short-range scattering becomes dominant. This results in a peak mobility between the two regimes (Figure 2a). The dependence of mobility on carrier concentration, $\mu(n)$, is affected by the density of impurities, $n_0$, and by their distance from the graphene, $d_{imp}$. Hence, both $\delta n$ and the low-carrier mobility, $\mu(n \approx n_{NP})$ depend on $n_0$ and $d_{imp}$ (Figure 2b).

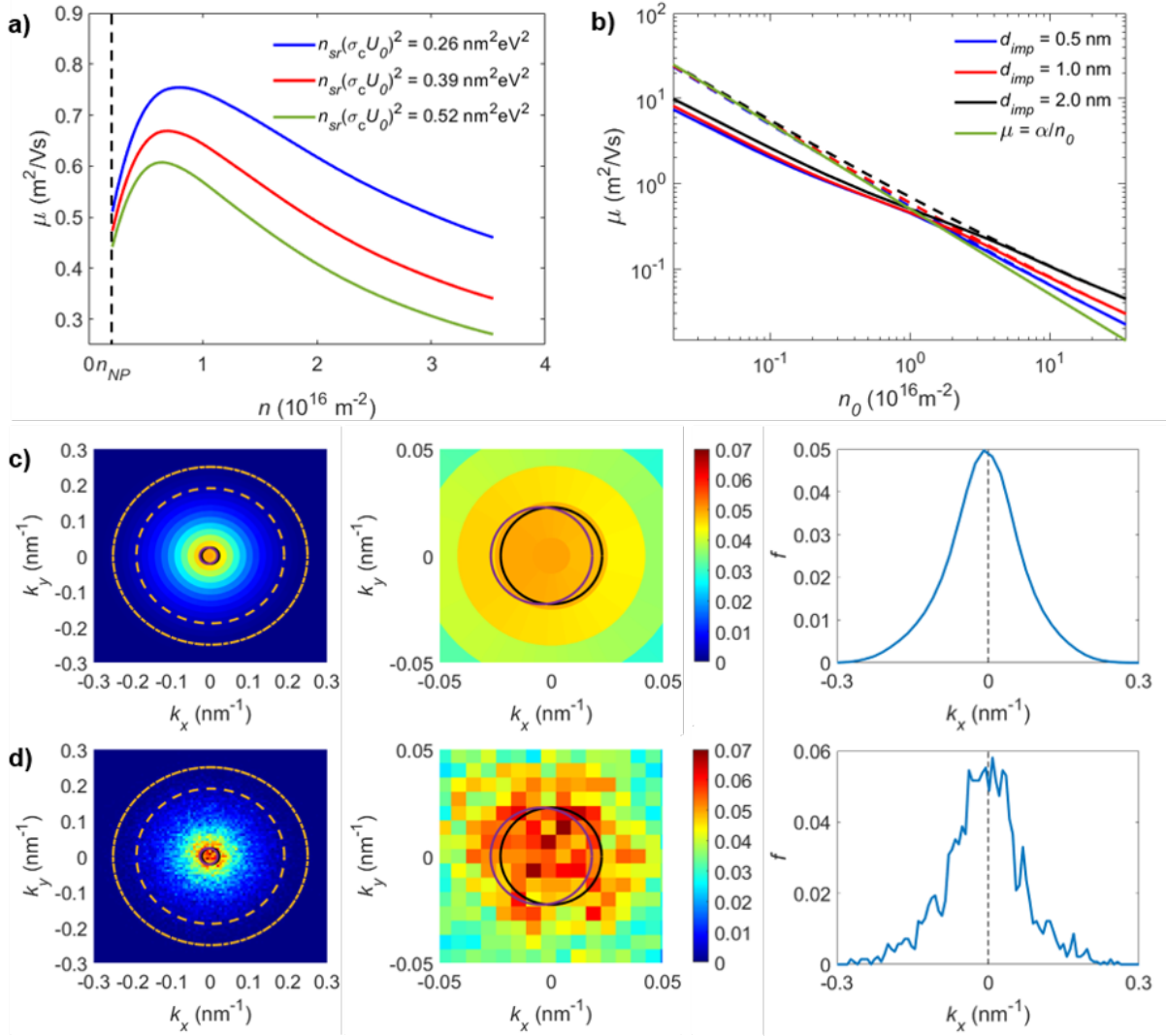

**Figure 2 (a)** Mobility, $\mu$, calculated using our DG model as a function of gate-induced carrier density, $n$, beyond the minimum carrier density, $n_{NP}$ (vertical dashed line), for varying short-range scattering strengths. The impurity density is $n_0 = 0.6 \times 10^{16}$ m$^{-2}$. **(b)** Mobility, $\mu$, calculated as a function of impurity density, $n_0$, for several stand-off distances, $d_{imp}$. Dashed curves are calculated using the linearised Boltzmann approximation, using Equation (6), the solid green curve shows the constant-capacitance approximation from Ref. [15], where $\alpha = 20e/h$. **(c)-(d)** Final distribution of electron momentum calculated for a small impurity density, $n_0 = 0.025 \times 10^{16}$ m$^{-2}$, at a distance $d_{imp} = 1$ nm and carrier density, $n = n_{NP} = 0.016 \times 10^{16}$ m$^{-2}$ found using (c) DG simulations and (d) Monte Carlo simulations. Black circle represents the initial Fermi circle, of radius $k_F$, purple circle shows a linear shift of the Fermi circle by $\delta k_x = eE\tau(\varepsilon_F)/\hbar$, orange dashed circle represents the $K$-phonon level, and the orange dashed-dotted circle represents the $\Gamma$-phonon level.

We see that the mobility increases as the graphene-impurity distance is increased. Furthermore, we observe that, for low impurity densities, the mobility given by the DG simulations differs from that obtained from the linearised Boltzmann calculations, whereas the two methods give $\mu$ values that converge at higher impurity densities. Both the linearised model and the DG simulations assume that initially, at $t = 0$, the electron gas is in thermal equilibrium and obeys the Fermi-Dirac distribution. Equation (9) assumes a linear shift in the momentum of the ensemble, proportional to the ensemble relaxation rate, $\tau(\varepsilon_F)$, whereas the DG simulations include the time evolution of the momentum distribution, described by the full Boltzmann Transport Equation (6). Therefore, the discrepancy between the two methods seen at low impurity densities can be explained by comparing the steady-state distribution functions. Figure 2c shows the final distribution of electrons obtained for a small impurity density, using the DG simulation. We obtain similar results by Monte Carlo simulations[28,37] (Figure 2d). In both the DG and Monte Carlo simulations, with increasing $t$ we observe continuous spreading of the electron distribution in $k$-space, until the occupied $k$ values become limited by inelastic phonon scattering. At this point the distribution is significantly different from the approximation of a linearly shifted Fermi-Dirac distribution function assumed in the derivation of Equation (9).

The effect of $d_{imp}$ on the electrical properties of graphene is summarised in Figure 3. Our calculations demonstrate that the linewidth of the $\rho(V_g)$ curves broadens with decreasing $d_{imp}$ (Figure 3a), which is accompanied by a decrease of carrier mobility due to enhanced charge scattering (Figure 3b). At small values of δ$n$, and hence small impurity densities, we observe discrepancy between the DG and linearised Boltzmann calculations of $\mu$. (compare Figure 2b and Figure 3b). Despite the discrepancy in $\mu(n_0)$, we find that both methods yield a similar δ$n(n_0)$ profile. In order to confirm these results, we now compare our calculations to experimental measurements.

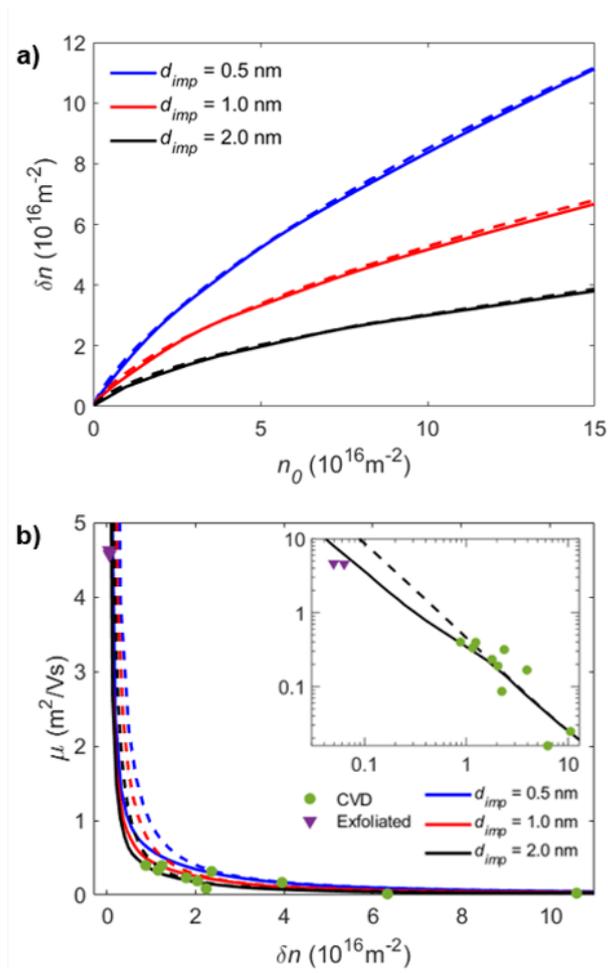

**Figure 3 (a)** FWHM, $\delta n$, variation with impurity density, $n_0$, for different stand-off distances $d_{imp}$ = 0.5 nm (blue curves), 1.0 nm (red curves), 2.0 nm (black curves). **(b)** Calculated mobility, $\mu$, versus FWHM, $\delta n$, curves (dashed and solid curves) compared to data from pristine graphene samples grown by CVD (filled circles) or exfoliated (filled triangles). The dashed lines are obtained from Equation (9), the solid lines are obtained using the DG simulations.

3. EXPERIMENTAL RESULTS

We apply our analysis to experimental results reported previously for >20 devices fabricated using exfoliated and CVD-grown graphene, and for heterostructures incorporating 2D (InSe, hBN) or 0D (colloidal QDs, inorganic perovskites) layers[38-40]. We fit the measured $\sigma(V_g)$ dependencies and determine values of $\mu$ and $\delta n$ (see Supplementary Information, SI1-2). In graphene devices, impurities at a distance, $d_{imp}$, from the 2D plane act as scattering centres. An increase in mobility is observed with decreasing δ$n$ (Figure 3a-b). These measurements are in good agreement with the results of our DG simulations with $d_{imp}$ = 2 nm. We note that our fit (Figure 3b) uses δ$n$ calculated from $\sigma_{FWHM}$, rather than $n_0$ extracted from the gate voltage at which $\sigma(V_g) = \sigma_{max}$. By using our simple expression for the conductivity, and assuming the universal minimum conductivity for pristine graphene, $\sigma_{min} \approx 4e^2/h$ [22], and constant mobility (with respect to carrier density), we obtain a simple inverse power law for mobility, which provides excellent agreement with experimental data from a range of devices (see Supplementary Information, SI2).

$$\mu = \frac{4\sigma_{min}}{e\delta n} = \frac{16e}{h\delta n}. \tag{20}$$

Recently, the decoration of graphene devices with other low-dimensional materials, such as 0D (colloidal PbS quantum dots[38] or $CsPbI_3$ perovskite[39]) and 2D (InSe flakes)[40] materials has been used to functionalise these devices, e.g. for photon sensing.[5,39,40] The properties of the graphene heterostructures are greatly affected by both the unintentional presence of charge impurities in the vicinity of graphene (as described above by $d_{imp}$) and those deliberately introduced by the top layer ($d_{top}$) in graphene heterostructures (Figure 4a), which we model as a distribution of impurities at some effective distance, $d_{eff}$. We note that in surface-decorated graphene devices, the distance between the graphene plane and the top layer can be controlled, for example by introducing a dielectric layer such as hBN, thus providing a tool for tailoring the electrical properties.

The relationship between mobility and the gate-voltage offset is $\mu \propto 1/n_0$ for most pristine devices[15]. However, for devices with high densities of correlated unipolar charges[32,33] or uncorrelated bipolar charges,[41] spatial correlation between charges must be considered. This is particularly important when the dopants are mobile and able to adopt low energy, correlated, configurations. Such effects were recently demonstrated for quantum dot-decorated graphene

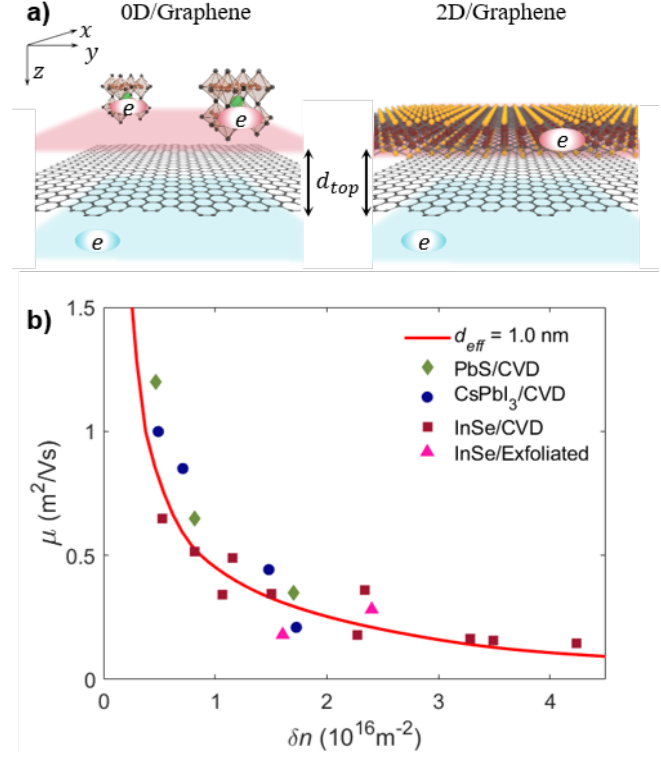

**Figure 4 (a)** Schematic diagram showing the position of impurities and surface-charges, due to 0D structures (e.g. perovskites) and 2D structures (e.g. InSe), with respect to the graphene sheet. **(b)** Relationship between mobility, $\mu$, and the FWHM, $\delta n$, obtained using the DG simulations, taking $d_{imp}$ = 1.0 nm, compared to data from multiple modified graphene samples (data points for each sample type are labelled as shown in the inset legend).

and validated using Monte Carlo simulations.[33,41] Despite the different $\mu(n_0)$ characteristics of decorated and pristine graphene, remarkably we find that both types of device exhibit the universal scaling behaviour that we have identified (Figure 4a). All our surface-decorated devices follow the trend given by our universal scaling model for pristine graphene with different impurity densities. The experimental results for the InSe, perovskite and PbS decorated SLG are best fitted by DG calculations when $d_{eff}$ = 1 nm. Therefore, we find that the relationship between $\mu$ and $\delta n$ is consistent throughout all of the devices, as can be expected from the analytical expression given in Equation (17), with modifications to only the effective distance of the impurities.

Our model links together three key transport parameters of SLG devices: $\mu$, $n_0$ and $\delta n$, where $\delta n$ can also be used to calculate $\rho_{max}$ and $\sigma(V_g)$ (see Supplementary Information SI3). Remarkably, this model can be used to extrapolate the whole $R(V_g)$ dependence from a single $R(V_g) = R_{max}$ measurement and for a wide range of different graphene devices (see Figure 1c-d). Of particular interest is the applicability of our model to a wide variety of different graphene types and to different device structures and geometries. Consequently, the model has the potential to both predict and explain the observed behaviour of newly emerging device concepts and graphene types. Recently, the need for upscaling of graphene growth and device manufacturing has led to significant research focus on Molecular Beam Epitaxial growth,[42] liquid exfoliation of 2D materials[43] and additive manufacturing (3D printing) of graphene devices.[44,45] Our preliminary results indicate that our model can be optimised and expanded to explain and predict the properties of 3D printed graphene devices, by accounting for flake-to-flake hopping of charges.

4. CONCLUSIONS

We have developed a universal analytical convolution model of electron transport in graphene and graphene heterostructures, supported by numerical time-dependent analysis of the charge carrier distributions. Our model includes the effects of impurities and optical phonons on the observed charge transport properties of graphene devices. We find that the properties of a wide range of devices, from high-quality graphene with low carrier density to graphene-based heterostructures, exhibit universal behaviour that can be accurately described with this model. Our calculations combine multiple parameters that affect charge transport in graphene, and facilitate the design, accurate ab-initio prediction of key transport parameters, and analysis of future electronic and opto-electronic devices based on 2D materials. Furthermore, our results may be generalised to predict and improve the electrical behaviour of 2D multilayers made by 3D printing, or from reduced graphene oxide, which are promising candidates for the scalable high-yield manufacture of large-area optoelectronic devices that harness the unique properties of 2D materials.

**Acknowledgements**

This work was funded by the Engineering and Physical Sciences Research Council [grant number EP/P031684/1]. Authors acknowledge support from Dstl and the EU Graphene Flagship.

**Author contributions**

JG and OM performed modelling studies, NC analysed experimental data. All authors contributed to design of the study, analysis and interpretation of results. All authors co-wrote and approved the manuscript.

**Competing Interests**

Authors declare no competing interests.

# SUPPLEMENTARY INFORMATION

# Universal mobility characteristics of graphene originating from electron/hole scattering by ionised impurities

Jonathan H. Gosling,[1-2] Oleg Makarovsky,[1] Feiran Wang,[2] Nathan D Cottam,[1]

Mark T. Greenaway,[3] Amalia Patanè,[1] Christopher J. Tuck,[2]

Lyudmila Turyanska[2*] and T. Mark Fromhold[1*]

[1]*School of Physics and Astronomy, University of Nottingham, Nottingham, NG7 2RD, UK.*

[2]*Centre for Additive Manufacturing, Faculty of Engineering, University of Nottingham, Nottingham, NG7 2RD, UK.*

[3]*Department of Physics, Loughborough University, Loughborough, UK*

## SI1. ANALYTICAL MODELLING OF SINGLE-LAYER GRAPHENE AND FITTING OF THE EXPERIMENTAL DATA.

Efficient gating provided by the Si/SiO$_2$ substrates allows easy measurement of the carrier concentration and mobility of graphene Field Effect Transistors (FETs), without requiring sophisticated equipment or an applied magnetic field. The carrier concentration of graphene can be estimated using a parallel plate capacitance model where the bottom plate is a Si substrate and top plate is the graphene layer. Then the capacitance of the graphene FET per unit area is $C_0 = \epsilon\epsilon_0/d = 115$ μF/m$^2$, where $d = 300$ nm is SiO$_2$ layer thickness and $\epsilon = 3.9$ is its dielectric constant. The parallel capacitance model of a graphene FET is based on the assumption that all mobile charges are equally distributed between the Si gate electrode and the graphene layer. In this case, the carrier number density in the graphene layer is:

$$n_c = |n_0 + \frac{C}{e}V_g|, \tag{S1}$$

where $n_0$ is the doping of the graphene when $V_g = 0$, e.g. the doping arising from the nearby charged impurities, and $V_g$ is the applied gate voltage. The doping level, $n_0$, is found from the position of the minimum, $\sigma_{min}$, (i.e. the maximum of the resistivity, $\rho_{max}$) in the $\sigma(V_g)$ curve. The linear relation, $n_c(V_g)$, defines the field effect mobility as $\mu_{FE} = \Delta\sigma/e\Delta n$, where $\Delta\sigma$ and $\Delta n$ are typically measured from the positions of the maximum gradients in the $\sigma(V_g)$ curve. The mobility of charge carriers in the graphene is assumed to be independent of $V_g$ [S1].

The two main assumptions introduced above, namely the parallel plate capacitor carrier concentration (Equation (S1)) and $V_g$-independent mobility, lead to a simple linear approximation of $\sigma(V_g)$ using the Drude formula:

$$\sigma(V_g) = e\mu n_c(V_g), \tag{S2}$$

where $n_c(V_g)$ is given by Equation (S1) and the mobility, $\mu$, is constant. This approach provides good estimates for the values of $n_0$ and $\mu_{FE}$, but does not explain several features of the experimental data, such as: (i) the sublinear $\sigma(V_g)$ variation far from the Dirac (minimum conductivity) point, (ii) the finite conductivity, $\sigma_{min} > 0$, when the chemical potential coincides with the Dirac point and (iii) the left-right asymmetry of the $\sigma(V_g)$ curve with respect to the $\sigma = \sigma_{min}$ ($n(V_g) = n_0$) point.

Sublinear $\sigma(V_g)$ variation has previously been explained as resulting from an interplay of short- and long-range scattering mechanisms. Short-range scattering can be described by introducing an additional constant $V_g$-independent resistivity, $\rho_s$, into the Drude formula [S1]:

$$\frac{1}{\sigma(V_g)} = \frac{1}{en_c(V_g)\mu} + \frac{1}{\rho_s}. \tag{S3}$$

The blue curves in Figures 1c-d of the main text show this fit taking $\rho_s = 60$ Ohm/sqr.

The finite value of the conductivity minimum, $\sigma_{min}$ (resistivity maximum, $\rho_{max}$) measured at the Dirac point ($n = n_0$ Figure 1c-d) has been widely considered [S1, S2, S5]. Graphene conductivity in the vicinity of the Dirac point is strongly affected by many different phenomena, such as quantum capacitance [S3-S4] and temperature [S5]. As a result, the classical parallel plate capacitor model is not applicable near the Dirac point and the carrier concentration never reaches zero, because some electrons and holes are present in the graphene even when the Fermi energy, $\varepsilon_F$, is at the Dirac point. This effect can be described as spatial fluctuations of the local electrostatic potential along the graphene layer and resulting electron and hole "puddles". As both electrons and holes play equal roles in the graphene conductivity, with no scattering on the borders between the n- and p-type graphene areas [S6], we define a minimum cumulative (electrons and holes) carrier concentration, $\delta n$.

Using experimental data, the minimum carrier concentration $\delta n$ is defined as a full width half maximum of the $\rho(n)$ peak, $\delta n_{FWHM}$. It can also be calculated by applying the Drude conductivity formula at the Dirac point so that $n_{NP} = \sigma_{min}/e\mu$. To expand this fit away from the Dirac point, the experimental data are fitted by a convolution of linear conductivity (Equation S2), or linear conductivity with a short-range scattering correction (Equation S3) and a carrier distribution function $f(n)$ (box function):

$$f(n) = \begin{cases} \frac{1}{\delta n} & \text{for } -\frac{\delta n}{2} < n < +\frac{\delta n}{2} \\ 0 & \text{for } -\frac{\delta n}{2} > n \text{ or } n > +\frac{\delta n}{2} \end{cases}, \quad (S4)$$

We note, that carrier distribution function can be also represented as $f(V_g)$, i.e. $n(V_g) = |\frac{CV_g}{e} + n_0| * f(V_g)$, and can be written as

$$f(V_g) = \begin{cases} \frac{1}{\delta n} & \text{for } |\frac{CV_g}{e} + n_0| < \frac{\delta n}{2} \\ 0 & \text{for } |\frac{CV_g}{e} + n_0| > \frac{\delta n}{2} \end{cases}. \quad (S5)$$

Therefore, we obtain

$$n(V_g) = \begin{cases} \left(\dfrac{\delta n}{4} + \dfrac{\left(n_0 + \dfrac{C_0}{e}V_g\right)^2}{\delta n}\right) & \text{for } \left|\dfrac{CV_g}{e} + n_0\right| < \dfrac{\delta n}{2} \\ \left|\dfrac{CV_g}{e} + n_0\right| & \text{for } \left|\dfrac{CV_g}{e} + n_0\right| > \dfrac{\delta n}{2} \end{cases}, \tag{S6}$$

from which we obtain the full gate-voltage dependence of the conductivity, $\sigma_f(V_g)=e\mu n(V_g)$. Convolution of the functions $y(x)=|x|$ (mimicking $\sigma(n)$ dependence (S2)) and $f(x)$ is presented in the supplementary video (convolution2.avi).

This convolution fit applied to the experimental data is shown by the red curves in Figures 1c-d of the main text. We used $\delta n$ as a fitting parameter and found that the best fit is obtained with $\delta n$ equal to $\delta n_{FWHM}$ (or $4n_{NP}$).

### SI2. PHENOMENOLOGICAL FIT OF THE $\mu(\delta n_{FWHM})$ DEPENDENCE

In this study, we analysed the data recorded in our previous work on graphene devices fabricated using single-layer graphene placed on ~300nm thick $SiO_2$/Si or on a few monolayer thick hBN/$SiO_2$/Si substrates with a bottom gate electrode. We used two device geometries: a 2-terminal diode and a Hall bar. Results obtained for the following devices are analysed in this work: pristine exfoliated single-layer graphene [S7]; pristine CVD-grown single-layer graphene (Graphenea and Graphene Supermarket) [S8]; graphene covered by a thin (from ~5 nm to ~50nm) layer of exfoliated InSe [S8]; graphene covered by a layer of inorganic perovskite $CsPbI_3$ nanocrystals [S9]; graphene covered by a layer of colloidal PbS quantum dots [S10].

The results of our electrical measurements were analysed to determine the underlying $\mu(\delta n_{FWHM})$ dependence. Assuming that away from the Dirac point, conductivity increases linearly with carrier density (i.e. constant mobility), it can be shown that

$$\mu = \dfrac{4}{e\rho_{max}}\dfrac{1}{\delta n}. \tag{S7}$$

For most pristine graphene samples, we can approximate $\rho_{max} \approx h/4e^2$ [S11], as shown in figure S1a. Given Equation (S7), this results in $\mu \approx 16e/h\delta n$ (see Figure S1b) for pristine samples. We note that for surface modified devices, the maximum conductivity is slightly larger, as shown in Figure S2a. This can be due to the increased amount of residual charge, as modelled by the smaller effective distance (Figures 3b and 4b of main text). If we approximate $\rho_{max} \approx h/5e^2$, we find $\mu \approx 20e/h\delta n$ for modified graphene (see Figure S2b). The simultaneous increase of mobility and impurity density has been previously explained using spatial correlation of charged impurities [S10].

Our results suggest that the characteristic carrier density fluctuation determined from the FWHM, $\delta n$, of the $R(V_g)$ dependence can be used to estimate the mobility of a wide range of graphene devices in way that is strikingly similar to the phenomenological relation defined for impurity density, i.e. $\mu = C/n_{imp}$ [S2].

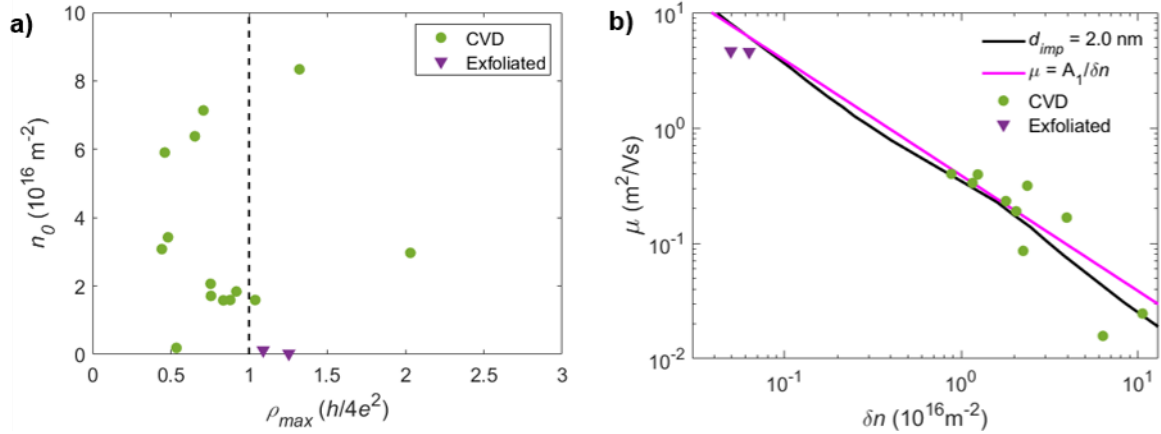

**Figure S1 (a)** Values of the maximum resistivity, $\rho_{max}$, measured for pristine graphene with different levels of doping, $n_0$. **(b)** Variation of the mobility, $\mu$, with the carrier density fluctuation, $\delta n$, measured for pristine graphene. Black curve shows our DG simulation, pink curve shows the analysis using equation S5, taking $A_1 = 16e/h\delta n$.

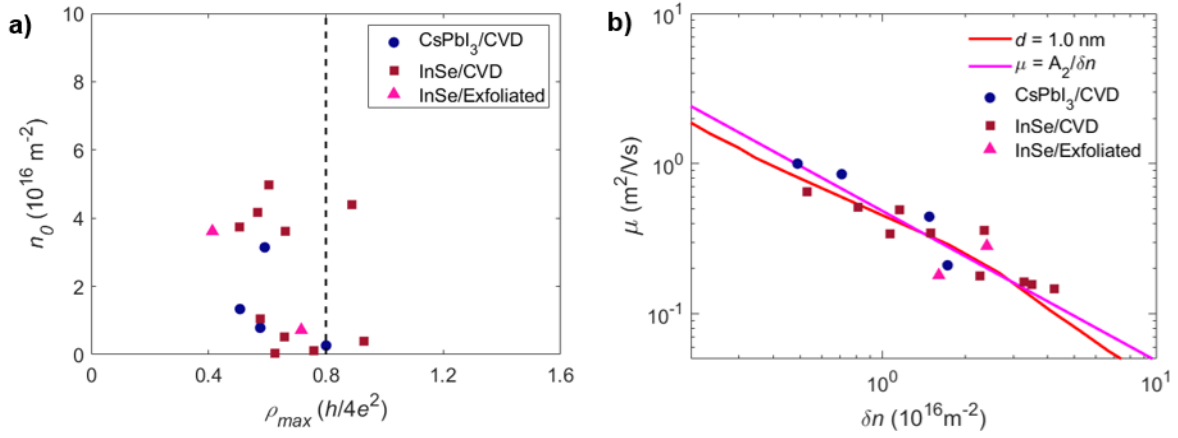

**Figure S2 (a)** Values of the maximum resistivity, $\rho_{max}$, measured for surface modified graphene with different levels of doping, $n_0$. **(b)** Variation of the mobility, $\mu$, with the carrier density fluctuation, $\delta n$, measured for pristine graphene. Red curve shows our DG simulation, pink curve shows the analysis using equation S5, taking $A_2 = 20e/h\delta n$.

## SI3. GOLDEN TRIANGLE OF GRAPHENE CONDUCTIVITY

Different phenomenological equations for graphene transport parameters used in our work can be represented graphically as the "Golden Triangle" shown in Figure S3. The dashed lines represent simple phenomenological relations between the four main parameters relevant to the conductivity of graphene: the resistivity at the Dirac point, $\rho_{NP}$, the carrier mobility, $\mu$, the doping level, $n_0$, and the uncertainty of the carrier density $\delta n$. The carrier density uncertainty parameter $\delta n$, provides additional connections between the other 3 parameters.

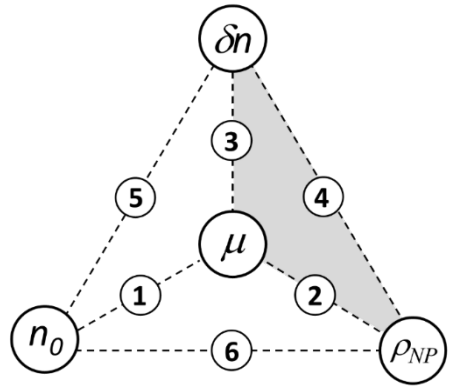

**Figure S3 (a)** Golden triangle of graphene conductivity. Phenomenological relations between the parameters at the vertices of the triangles are shown within numbered circles, corresponding to expressions (1-6) below.

Within the Golden triangle, the numbered circles along the sides of the triangles refer to expressions (1-6) below, which relate the parameters at the corresponding vertices. These expressions provide estimates of key graphene parameters from an incomplete data set as well as for semi-analytical fitting of the experimental data:

**(1)** $\mu = C_1/n_0$, where $C_1 = 5\times10^{15}$ V$^{-1}$s$^{-1}$ [S2] assuming that the graphene doping level equals the concentration of ionised impurities: $n_0 = n_{imp}$. This simple fit, our experimental data, and detailed numerical modelling is shown in Figure S4 a).

**(2-3-4)** The shaded part of the Golden triangle can be described by an analytical formula $\mu = \frac{4}{e\delta n \rho_{NP}}$, which is derived from the definition of the uncertainty of the carrier density due to local potential fluctuations $\delta n = \delta n_{FWHM}$, which is the measured FWHM of $\rho(n)$.

**(3)** $\mu$ **vs** $\delta n$ can be obtained from (5) and (1) taking $\delta n \approx 0.8\, n_0$ and $\mu \approx 5\times10^{15}/n_0$, which gives $\mu \approx 4\times10^{15}/\delta n$ (Figure S4d).

**(5)** $\delta n$ **vs** $n_0$ can be found from (6) and (2-3-4): $\rho_{NP} = \frac{h}{4e^2}$ and $\mu = \frac{4}{e\delta n \rho_{NP}}$. Then using (1) [$\mu = C_1/n_0$] we can find $\delta n = C_2 n_0$, where $C_2 \approx 0.8$ (Figure S4 b).

**(6)** $\rho_{NP}$ **vs** $n_0$: this relation is somewhat uncertain due to significant scatter in the experimental data (Figure S4c). Nevertheless, a simple quantum conductivity model with $\rho_{NP} = \frac{h}{4e^2} = 6453\,\Omega$ [S11], provides a reasonable fit to the experimental data.

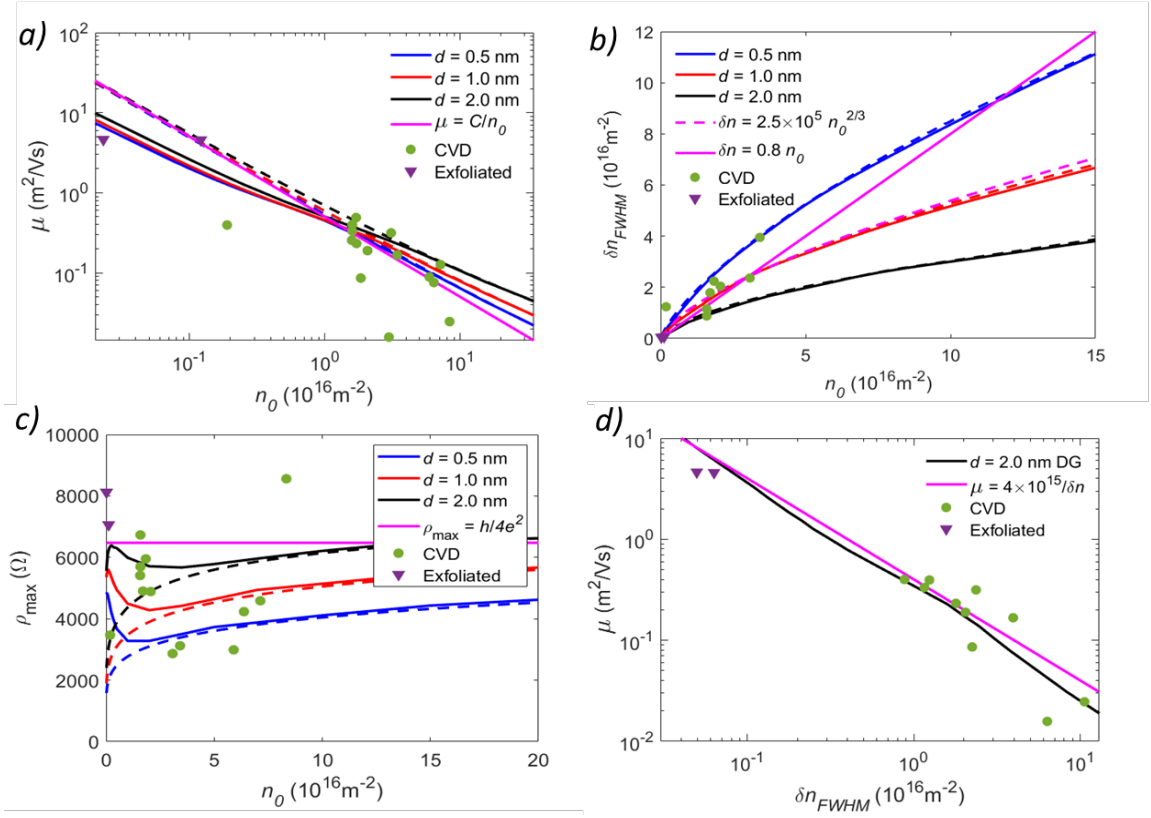

**Figure S4 (a-d)** Phenomenological equations (1-6) (magenta lines), our corresponding experimental data (symbols for the sample types shown in the legends) and full numerical calculations (red, black and blue curves).

Expressions (1-6) can be applied to simulate the $\sigma(V_g)$ and $R(V_g)$ dependences using only one input parameter: $n_0$, $\delta n$ or $\mu$. Results of such simulations for different $n_0$ are shown in Figure S5. This model is not perfect. For example, a large uncertainty comes from the constant maximum resistance $\rho_{NP} = \frac{h}{4e^2}$ (Expression (6)). In real devices, $\rho_{NP}$ tends to decrease with increasing $n_0$ and may also depend on temperature [S1]. Note that there is a single intersection point at $V_g=0$ for all of the $\sigma(V_g)$ curves and another for all the $R(V_g)$ curves. This is due to the simplified mobility (1): $\mu = \frac{C_1}{n_0}$ and because $n = n_0 + \frac{C}{e}V_g = n_0$ when $V_g = 0$. Consequently, using the Drude formula gives $\sigma(V_g = 0) = en_0\mu = eC_1$=8.011x10$^{-4}$ 1/Ohm, which is constant for any device with any $n_0$. This simple approximation has limited quantitative accuracy in real devices, see e.g. [S1, S5, S10].

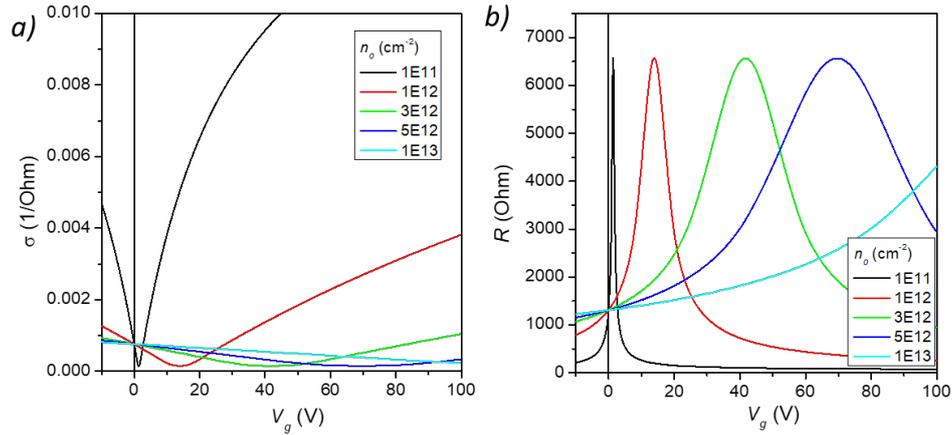

**Figure S5.** The $\sigma(V_g)$ **(a)** and $R(V_g)$ **(b)** dependences generated using Equations (1-6) for a range of $n_0$ values (see legends inset) assuming a conventional 300 nm $SiO_2$/Si substrate. Additional short range-scattering was introduced by setting $R_s = 60\ \Omega$.